\documentclass[12pt]{article}

\usepackage[sort&compress,numbers]{natbib}
\usepackage{amsmath}
\usepackage[utf8]{inputenc}
\usepackage{amsmath,amssymb,mathtools}
\usepackage{graphicx}
\usepackage[hidelinks, pageanchor=false]{hyperref}
\usepackage{xcolor}
\usepackage{multirow}
\usepackage{babel}
\usepackage{xspace}
\usepackage{enumerate}
\usepackage{caption}
\usepackage{subcaption}
\usepackage{soul}
\usepackage{float} 
\usepackage{comment}
\usepackage{longtable}
\usepackage[
  a4paper,
  twoside=false,
  left=1.7cm,
  right=1.7cm,
  top=2.5cm,
  bottom=3cm
]{geometry}

\renewcommand{\thefigure}{\arabic{figure}}  
\allowdisplaybreaks
 
\def\beq{\begin{equation}}
\def\eeq{\end{equation}}
\def\bea{\begin{align}}
\def\eea{\end{align}}

\def\Eq#1{Eq.~(\ref{#1})}

\begin{document} 
\begin{titlepage}
\renewcommand{\thefootnote}{\fnsymbol{footnote}}
\begin{flushright}
IFIC/23-16  \\  FTUV-23-0502.2201
\end{flushright}
\par \vspace{10mm}
\begin{center}
{\Large \bf
Quantum Fourier Iterative Amplitude Estimation\\
}
\end{center}
\par \vspace{2mm}
\begin{center}
{\bf Jorge J. Mart\'{\i}nez de Lejarza$^a$, Michele Grossi$^b$, Leandro Cieri$^a$ and Germ\'an Rodrigo$^a$}\\  
\vspace{5mm}

$^{(a)}$ 
Instituto de F\'{\i}sica Corpuscular, Universitat de Val\`encia - Consejo Superior de Investigaciones Cient\'{\i}ficas, Parc Cient\'{\i}fic, E-46980 Paterna, Valencia, Spain

$^{(b)}$ European Organization for Nuclear Research (CERN), 1211 Geneva, Switzerland

\vspace{5mm}

\end{center}

\par \vspace{2mm}
\begin{center} {\large \bf Abstract} \end{center}
\pretolerance 10000

\indent Monte Carlo integration is a widely used numerical method for approximating integrals, which is often computationally expensive. In recent years, quantum computing has shown promise for speeding up Monte Carlo integration, and several quantum algorithms have been proposed to achieve this goal. 

In this paper, we present an application of Quantum Machine Learning (QML) and Grover's amplification algorithm to build a new tool for estimating Monte Carlo integrals. Our method, which we call Quantum Fourier Iterative Amplitude Estimation (QFIAE), decomposes the target function into its Fourier series using a Parametrized Quantum Circuit (PQC), specifically a Quantum Neural Network (QNN), and then integrates each trigonometric component using Iterative Quantum Amplitude Estimation (IQAE). This approach builds on Fourier Quantum Monte Carlo Integration (FQMCI) method, which also decomposes the target function into its Fourier series, but QFIAE avoids the need for numerical integration of Fourier coefficients. This approach reduces the computational load while maintaining the quadratic speedup achieved by IQAE.

To evaluate the performance of QFIAE, we apply it to a test function that corresponds with a particle physics scattering process and compare its accuracy with other quantum integration methods and the analytic result. Our results show that QFIAE achieves comparable accuracy while being suitable for execution on real hardware. We also demonstrate how the accuracy of QFIAE improves by increasing the number of terms in the Fourier series.

\end{titlepage}
\section{Introduction}
The quantum amplitude estimation (QAE) method~\cite{Brassard_2002} is a quantum algorithm used to estimate the value of a certain amplitude of a specific quantum state. QAE uses amplitude amplification, which is a generalization of Grover's searching algorithm~\cite{Grover:1997fa}, to increase the probability of measuring the desired state (good state) over the non-desired state (bad state).
This algorithm has been proposed for a wide range of problems, such as solving large linear systems of equations~\cite{Harrow_2009}, optimization~\cite{Kaneko2021} and numerical integration~\cite{Montanaro2015, Pooja}, which is the application that concerns us. The main problem of the original QAE algorithm~\cite{Brassard_2002} is that one of its key components is a subroutine called Quantum Phase Estimation (QPE)~\cite{9781107002173}, which is composed of many operations considered expensive for the current Noisy Intermediate Scale Quantum (NISQ) devices. Hence, the application of QPE could potentially suppress the speedup that Grover's algorithm provides. 

Recently, various solutions have been suggested to tackle this issue, see~\cite{Intallura:2023yvu} for a detailed review of different proposals. In~\cite{Suzuki2020}, the authors propose using a set of Grover iterations along with Maximum Likelihood Estimation (MLE), which is named Maximum Likelihood Amplitude Estimation (MLAE). In~\cite{Plekhanov:2021kir}, this method was extended by using Variational Quantum Algorithm (VQA) techniques in an attempt to decrease the Grover iterations by implementing variational steps. Another study~\cite{wie2019simpler} suggests replacing QPE with the Hadamard test, which is similar to Kitaev's Iterative QPE~\cite{Kitaev2002ClassicalAQ}. While the mentioned works offer potential simplifications of QAE, they do not provide rigorous proofs of the rightness of their proposed algorithms. On the other hand, in~\cite{aaronson} a variant is proposed, that, unlike the others, proved rigorously that QAE without QPE can achieve a quadratic speedup over classical Monte Carlo (MC) simulation. However, the number of involved constants is very large and likely to make the algorithm impractical unless further optimized. A more recent study~\cite{Ghosh:2023qze} introduces a novel version that maintains the quantum advantage by employing dynamical circuits. This approach involves integrating classical processing within the coherence time of the qubits. Nevertheless, it is noteworthy that this algorithm may rely on specific hardware and is currently exclusive to IBM quantum~\cite{dynamicIBM}.

In~\cite{Grinko_2021} another variant is proposed that combines the probably and desired asymptotic speedup with a constant factor that is shown to be numerically small in most of the cases. This new version is called Iterative Quantum Amplitude Estimation (IQAE) and it is the variant of the algorithm that will be used as a reference in the following.

Monte Carlo integration is a numerical method for approximating definite integrals using random sampling. It works by generating a large number of random samples from the region of integration and then using the average value of the function evaluated at these samples as an estimate of the true integral value. This method is particularly useful for high-dimensional integrals and is extensively used in particle physics~\cite{Hahn:2004fe} and many other fields.
Nevertheless, this method could be very computationally demanding for certain kinds of integrals. In this context, a Quantum Monte Carlo Integration (QMCI) method that achieves a quadratic speedup over the number of queries with respect to its classical counterpart by using QAE has been presented in~\cite{Montanaro2015}. However, this proposal does not deal properly with the problem of how to prepare the initial quantum state, and in some cases, it involves a lot of arithmetic to be performed quantumly which could potentially wipe out the quantum advantage.

In~\cite{Herbert_2022}, a new approach to QMCI is introduced that takes full advantage of the power of quantum computing without the need for arithmetic or phase estimation on the quantum computer. This method is unique as it achieves all these aspects at once, which was not seen before in previous proposals. The central idea of the method introduced there is using Fourier series decomposition to approximate the integrand in Monte Carlo integration, and then estimating each component individually through quantum amplitude estimation.

Nonetheless, the Fourier Quantum Monte Carlo Integration~(FQMCI) method also has some drawbacks, as discussed in Sec.\ref{sec:fourierintegration}, as it does not take into account the computation of the Fourier series itself. In this paper, we propose a novel approach to address this issue. Specifically, we suggest, and implement for a benchmark example, a quantum neural network (QNN) to fit the desired function and then extract the corresponding Fourier series that represents the quantum circuit. This enables an end-to-end quantum algorithm that might retain the quadratic speedup in Monte Carlo integration.

\section{Iterative Quantum Amplitude Estimation}
\label{sec:qae}

The Iterative Quantum Amplitude Estimation~(IQAE) algorithm proposed in~\cite{Grinko_2021} is a variant of the QAE algorithm which relies only on Grover's algorithm. This version replaces the Quantum Phase Estimation~(QPE) subroutine for a classical efficient post-processing method to reduce the required number of qubits and gates.
The IQAE algorithm consists of two components:
quantum amplitude amplification and an iterative search of the optimal number $k$ of amplifications that have to be done to succeed in the estimation.
The amplitude amplification~\cite{Brassard,PhysRevLett.80.4329,Brassard_2002} algorithm, being the generalization of the well-known Grover's quantum search algorithm~\cite{10.1145/237814.237866}, also achieves a quadratic speedup over its classical analog.

Let us consider a unitary operator $\mathcal{A}$ acting on $n+1$ qubits in such a way that:
\begin{equation}
    |\psi\rangle=\mathcal{A}|0\rangle_{n+1}=\sqrt{a}|\tilde{\psi}_1\rangle|1\rangle +\sqrt{1-a}|\tilde{\psi}_0\rangle|0\rangle,
    \label{eq:A}
\end{equation}
where $a\in[0,1]$ is the parameter to be estimated, while $|\tilde{\psi}_1\rangle$ and $|\tilde{\psi}_0\rangle$ are the $n$-qubit good and bad states, respectively. Using~\Eq{eq:A}, the parameter $a$ can be inferred from the ratio of obtaining the good and bad states. However, this does not provide any speedup, since its query complexity would be the same as the classical one.

The quantum advantage is achieved through the use of amplitude amplification, which allows for a quadratic speedup in the number of queries by applying the amplification operator:
\begin{equation}
    \mathcal{Q}=-\mathcal{A}S_0\mathcal{A}^{-1}S_{\chi},
    \label{eq:Q}
\end{equation}
where the operator $S_0$ tags with a negative sign the $|0\rangle_{n+1}$ state and does nothing to the other states, and the operator $S_\chi$ changes the sign to the good state, i.e. $S_\chi |\tilde{\psi}_1\rangle|1\rangle=-|\tilde{\psi}_1\rangle|1\rangle$.

By defining a parameter $\theta_a\in [0,\pi/2]$ such that $\sin^2{\theta_a}=a$, we have
\begin{equation}
    |\psi\rangle=\mathcal{A}|0\rangle_{n+1}=\sin{\theta_a}|\tilde{\psi}_1\rangle|1\rangle +\cos{\theta_a}|\tilde{\psi}_0\rangle|0\rangle.
    \label{eq:A2}
\end{equation}
Brassard et al. showed in~\cite{Brassard_2002} that repeatedly applying $\mathcal{Q}$ for $k$ times on $\psi$ results in
\begin{equation}
    \mathcal{Q}^k|\psi\rangle=\sin{((2k+1)\theta_a)}|\tilde{\psi}_1\rangle|1\rangle +\cos{((2k+1)\theta_a)}|\tilde{\psi}_0\rangle|0\rangle.
    \label{eq:Qk}
\end{equation}

\Eq{eq:Qk} represents that, after applying $k$ times
the operator $\mathcal{Q}$, we can obtain the good state with a probability of at least $4k^2$ times larger than that obtained from \Eq{eq:A2} for sufficiently small~$a$. This is the quadratic speedup obtained from amplitude amplification, since $2k$ measurements from $\mathcal{A} |0\rangle_{n+1}$, only give the good state with probability $2k$ times larger.
Finally, if we can infer the ratio of the good state after amplitude amplification, we can estimate the value of $a$ from the number of queries required to obtain such a ratio. 

The procedure to find the optimal number $k$ constitutes the original proposal from the IQAE algorithm. This subroutine considers a confidence interval $[\theta_l,\theta_u] \subseteq [0,\pi/2]$ with $\theta_l < \theta_a < \theta_u$, and a power $k$ of $\mathcal{Q}$ as well as an estimate for $\sin^2((2k+1)\theta_a)$. Then, applying the trigonometric identity $\sin^2(x)=(1-\cos(2x))/2$, our estimates for $\sin^2 ((2k + 1)\theta_a)$ are translated into estimates for $\cos((4k + 2)\theta_a)$ taking into account that we should know whether the argument is restricted to either $[0,\pi]$ or $[\pi, 2\pi]$. Therefore, we are interested in finding the largest $k$
such that the interval $[(4k+ 2)\theta_l,(4k+ 2)\theta_u]_{\mod 2\pi}$
is fully contained either in the upper or lower half-plane. This implies an upper bound of $k$, and the heart of this algorithm is the method used to find the next $k$ given $[\theta_l,\theta_u]$.


\section{Quantum Monte Carlo Integration}
\label{sec:montecarlo}

Monte Carlo integration is a very powerful tool employed in several fields such as particle physics, finance, or cosmology that uses random numbers to estimate a definite integral
\begin{equation}
     I=\int_{x_{min}}^{x_{max}} p(x)f(x) dx.
\end{equation}
The way to compute this integral using the Monte Carlo method is to choose $2n$ samples $x_i\in \{ 0,1\}^n$, with $i=0,\ldots, 2n-1$, independent and identically distributed (i.i.d.) from the probability identity function $p(x)$ in the interval $[x_{min},x_{max}]$. The integration value is approximated by the mean of $f(x)$ for all samples $x_i$.

The discrete Monte Carlo integration's main goal is to compute the expected value of a real function $0\leq f(x) \leq 1$ defined for $n-$bit input $x\in \{ 0,1\}^n$ with probability $p(x)$:
\begin{equation}
    \mathbb{E}[f(x)]=\sum_{x=0}^{2^n-1} p(x)f(x) .
\end{equation}

This computation might be extremely demanding for integrating some functions, and thus why a quantum version was developed~\cite{Montanaro2015}. 
The convergence of the Quantum Monte Carlo Integration (QMCI), measured in terms of the Mean Square Error, presents a quadratic advantage in terms of the number
of samples from the probability distribution. This quadratic speedup comes from the fact that it uses quantum amplitude estimation.

As explained in~\cite{Suzuki2020}, in a quantum algorithm for  Monte Carlo integration, the operator $\mathcal{A}$ is composed of two operators:
\begin{equation}
    \mathcal{P}|0\rangle_n=\sum_{x=0}^{2^n-1} \sqrt{p(x)}|x\rangle_n ,
\end{equation}
that encodes the probability function $p(x)$ into the state $|0\rangle_n$, and
\begin{equation}
    \mathcal{R}|x\rangle_n|0\rangle=|x\rangle_n \left(  \sqrt{f(x)}|1\rangle +\sqrt{1-f(x)}|0\rangle \right),
\end{equation}
which encodes the $f(x)$ function into an ancillary qubit that is added to the circuit. Therefore, applying the operator $\mathcal{A}$ to the $(n+1)$-qubit initial state one gets:
\beq
\begin{split}
|\psi\rangle &= \mathcal{A}|0\rangle_{n+1}=\mathcal{R}(\mathcal{P}\otimes\mathbb{I}^1)|0\rangle_{n+1}\\
&=\sum_{x=0}^{2^n-1} \sqrt{p(x)}|x\rangle_n \left( \sqrt{f(x)}|1\rangle +\sqrt{1-f(x)}|0\rangle \right),
\end{split}
\eeq
where $\mathbb{I}^1$ is the identity operator acting on the ancillary qubit. Now, we conveniently define the following quantities:
\begin{equation}
a=\sum_{x=0}^{2^n-1} p(x)f(x)= \mathbb{E}[f(x)],
\end{equation}
\begin{equation}
    |\tilde{\psi}_1\rangle=\frac{1}{\sqrt{a}}\sum_{x=0}^{2^n-1} \sqrt{p(x)}\sqrt{f(x)}|x\rangle_n,
\end{equation}
\begin{equation}
 |\tilde{\psi}_0\rangle=\frac{1}{\sqrt{1-a}}\sum_{x=0}^{2^n-1} \sqrt{p(x)}\sqrt{1-f(x)}|x\rangle_n,
\end{equation}
and the state $|\psi\rangle$ is rewritten as:
\begin{equation}
 |\psi\rangle=\sqrt{a}|\tilde{\psi}_1\rangle|1\rangle+\sqrt{1-a}|\tilde{\psi}_0\rangle|0\rangle,
 \label{eq:psi}
\end{equation}
which looks identical to \Eq{eq:A}. Then, the Monte Carlo integration can be regarded as amplitude estimation since the probability of the state, $a$, corresponds to the integral that has to be computed, $\mathbb{E}[f(x)]$. The only element that is left is the operator $\mathcal{Q}$ that can be computed using $U_{\psi}$ $U_{\psi_0}$:
\begin{equation}
\mathcal{Q}=U_{\psi}U_{\psi_0},  
\end{equation}
where
\begin{equation}
\begin{split}
& U_{\psi_0}=\mathbb{I}_{n+1}-  2\mathbb{I}_n|0\rangle \langle 0|, \,  \hspace{0.4cm} \, \\ & U_{\psi}=\mathbb{I}_{n+1}-2|\psi\rangle  \langle\psi|,  \qquad \mathbb{I}_n\equiv \mathbb{I}^{\otimes n}.
\end{split}
\end{equation}

\begin{figure}[ht!]
       \centering
  \includegraphics[width=.69\linewidth]{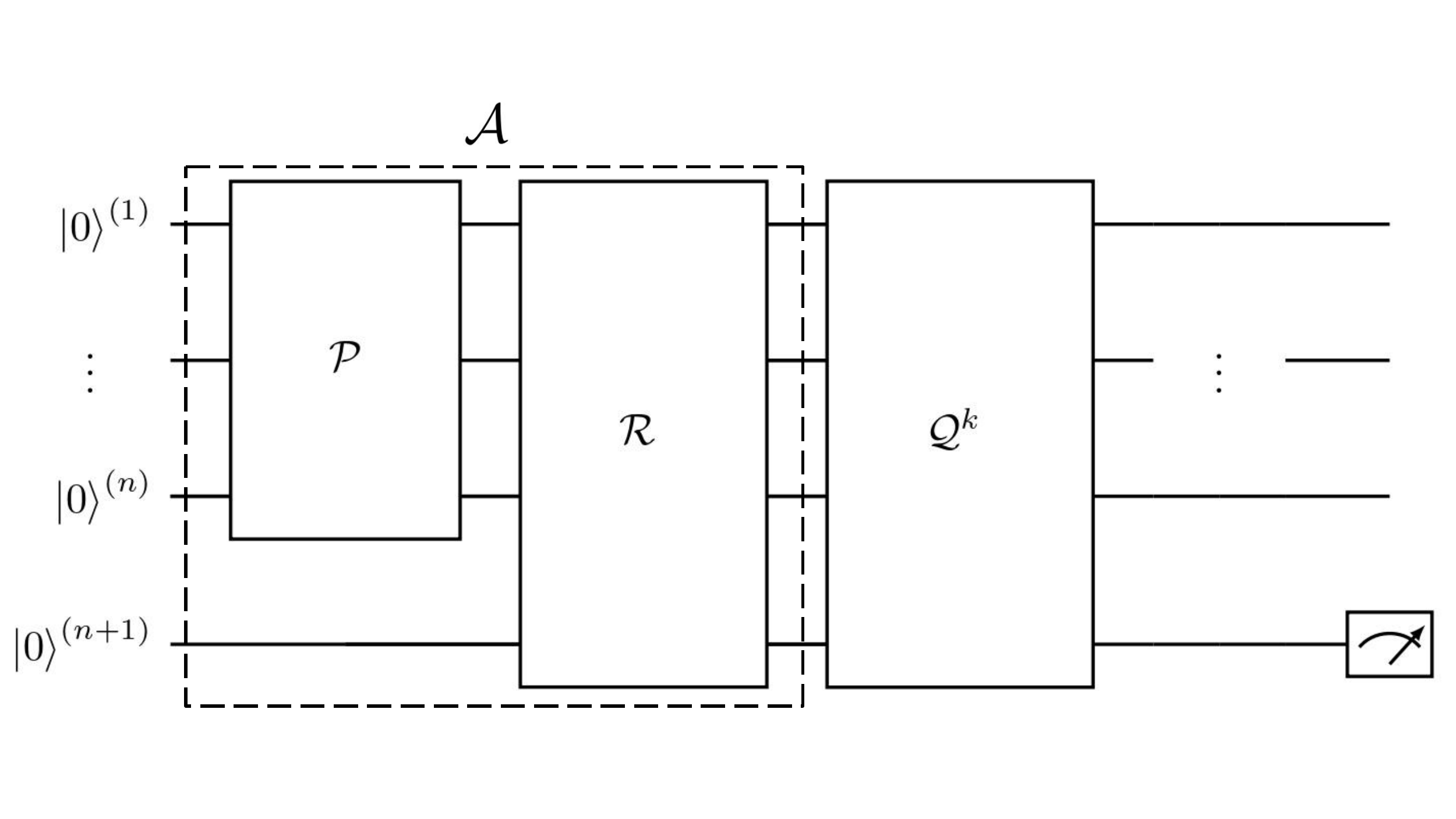}  

        \caption{Quantum circuit for amplitude amplification and estimation in Monte Carlo integration.}
        \label{fig:qae_circuit}
\end{figure}

Now, by putting $a=\sin^2{\theta_a}$, and using~\Eq{eq:psi} we are able to apply the IQAE algorithm previously described to Monte Carlo integration. The quantum circuit that represents this algorithm is shown in Fig. \ref{fig:qae_circuit}. 

\section{Fourier series integration}
\label{sec:fourierintegration}

As shown in Sec.~\ref{sec:montecarlo}, any function $f(x)$ that follows a probability distribution $p(x)$ could in principle be integrated using the QMCI method and this would yield a quantum speedup. Nevertheless, as pointed out in~\cite{Herbert_2022}, this comes at the cost of creating the circuit $\mathcal{R}$, which encodes the function $f(x)$, and in general, this will be prohibitively complex. Even the trivial
case where $f(x) = x$ (i.e., just finding the
mean of $p(x)$) would imply significant amounts
of arithmetic to be performed quantumly. However, an important exception is the specific case where $f(x) = \sin^2
(mx + c)$ for some constants $m$ and $c$, which can be encoded by a bank of $R_y$ rotation gates.

So, with this in mind, the author in~\cite{Herbert_2022} came up with the idea of decomposing the desired function $f(x)$ into sines and cosines. The procedure consists of extending $f(x)$ as a periodic piecewise function. Let us consider \textbf{f}$(x)$, which repeats with period $x_{\Tilde{u}}$--$ x_l$:
\begin{equation}
\textbf{f}(x)=
    \begin{cases}
         f(x) & \text{if } x_l\leq x \leq x_u \\
        \Tilde{f}(x) & \text{if } x_u\leq x \leq x_{\Tilde{u}},
    \end{cases}
    \label{eq:fx}
\end{equation}
where $\Tilde{f}(x)$ is itself sufficiently smooth and chosen such that $f(x_l) =\Tilde{f}(x_{\Tilde{u}})$ and $f(x_u) =\Tilde{f}(x_u)$, which holds true for their derivatives w.r.t to $x$. Since \textbf{f}$(x)$ is periodic, it has a Fourier series:
\begin{equation}
\textbf{f}(x)=c+ \sum_{n=1}^\infty \left( a_n\cos(n\omega x)+ b_n\sin(n\omega x) \right),
    \label{eq:ffourier}
\end{equation}
where $\omega = 2\pi/T$ and $T$ is the period of the periodic piecewise function. Those various trigonometric components of $\textbf{f}(x)$ can be estimated individually, using IQAE, and then recombined to get the full result.

Furthermore, it is shown that the quadratic quantum advantage is indeed retained with this method under certain conditions. 
Those include the ability to prepare a specific probability distribution as a shallow-depth quantum state and apply a function to random samples such that the mean cannot be calculated analytically. Additionally, it is important to be cautious when evaluating Fourier coefficients as numerical integration may shift the computational load rather than reduce the complexity. In~\cite{Herbert_2022}, the author claims that for commonly used functions such as the mean ($f(x)=x$), Fourier coefficients can be calculated symbolically. Moreover, he adds that if the coefficients cannot be found symbolically, for commonly used functions it may be reasonable to assume that they have been pre-computed and stored in advance. Therefore they would not add to the complexity of any individual run. Nevertheless, in the most general possible scenario, the Fourier coefficients need to be calculated and we cannot rely on them being symbolically computable or stored in advance. Hence, to attain the advertised speedup for a whole process, a new procedure to estimate the Fourier coefficients without relying on numerical integration must be introduced. This leads us to propose in Sec.~\ref{sec:qf} a novel method for obtaining the Fourier series of a function using a QNN.


\section{Quantum Fourier series}
\label{sec:qf}
The emerging field of quantum machine learning~(QML)~\cite{biamonte2017quantum} holds promise for enhancing the accuracy and speed of machine learning  algorithms by utilizing quantum computing~(QC) techniques. In the QML scenario, the quantum counterparts of classical neural networks, QNNs~\cite{abbas2021power}, have emerged as the de facto standard model for solving supervised and unsupervised learning tasks in the quantum domain.

The quest for quantum algorithms able to be executed on NISQ systems led to the concept of Variational Quantum Circuits (VQCs), i.e. quantum circuits based on a hybrid quantum-classical optimization framework,  namely hybrid algorithms that rely on a continuous interaction between a quantum computer and a classical computer.
VQCs are currently believed to be promising candidates to harness the potential of QC and achieve a quantum advantage~\cite{tilly2022variational,di2022quask,liu2021rigorous}.

\begin{figure}[ht!]
       \centering
  \includegraphics[width=.69\linewidth]{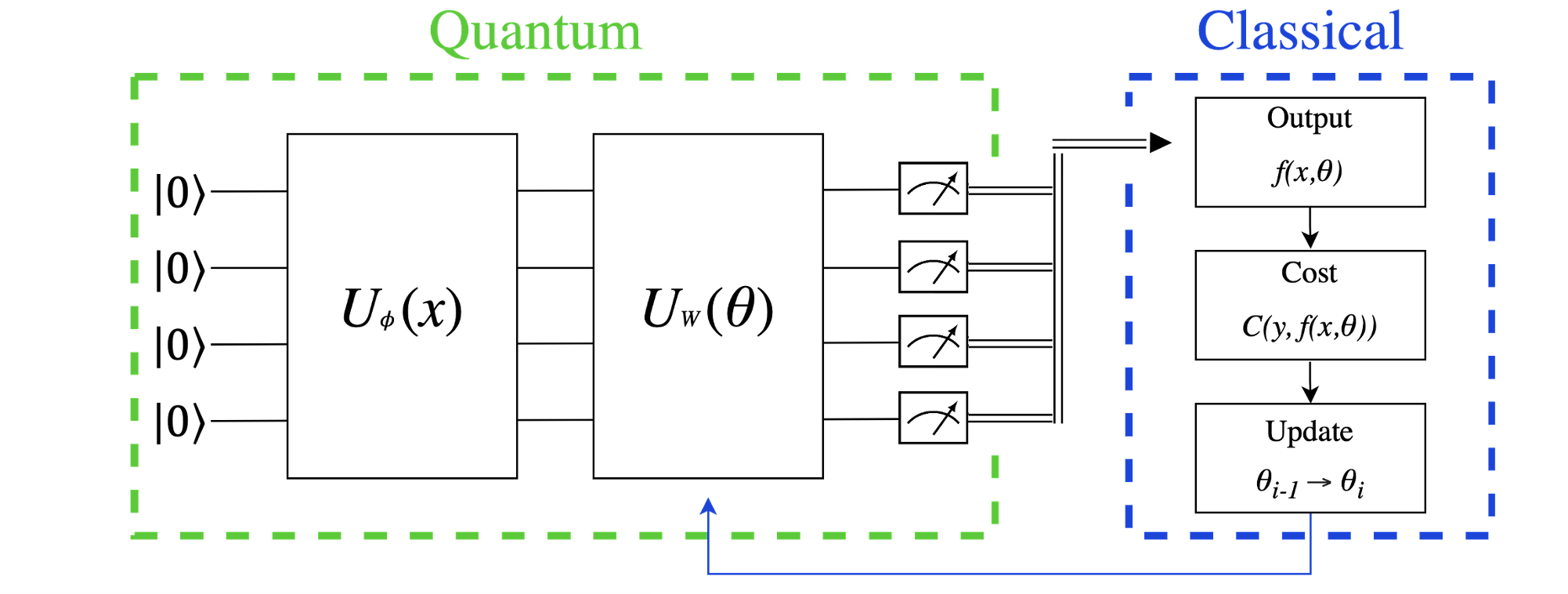}  

        \caption{Flowchart of a VQC. In the green square, there are 2 unitaries: a feature map $U_{\phi}(x)$ to encode the data, and the parametrized circuit $U_{w}(\theta)$ which is iteratively optimized with a classic routine, represented in the blue square. The optimization starts by obtaining the output of the quantum circuit and feeding the loss function. The updated parameters are then encoded back into the parametrized circuit. }%
        \label{fig:vqc_sketch}
\end{figure}

VQCs rely on a hybrid quantum-classical scheme~\cite{cerezo2021variational,mitarai_2018}, where an initial parameterized quantum circuit (PQC) is defined (called \textit{Ansatz}) and then using a classical optimizer this circuit is trained iteratively. The update of the parameters is driven by the evolution of the related loss function, as illustrated in Fig. \ref{fig:vqc_sketch}. 
This way, low-depth quantum circuits can be efficiently designed and implemented on the available NISQ devices; the noisy components of the quantum process are mitigated by the low number of quantum gates present in the VQCs.

VQCs are incredibly well-suited for the realization of QNNs with a constraint on the number of qubits~\cite{massoliALeap2022}. A QNN is usually composed of a layered architecture able to encode input data into quantum states and perform heavy manipulations in a high-dimensional feature space. The encoding strategy and the choice of the circuit Ansatz are critical for the achievement of superior performances over classical NNs: more complex data encoding with hard-to-simulate feature maps could lead to a concrete quantum advantage, but too expressive quantum circuits may exhibit flatter cost landscapes and result in untrainable models~\cite{holmes2022connecting}. 

\begin{figure}[ht!]
       \centering
  \begin{subfigure}[b]{0.5\textwidth}
  
  \includegraphics[width=.99\linewidth]{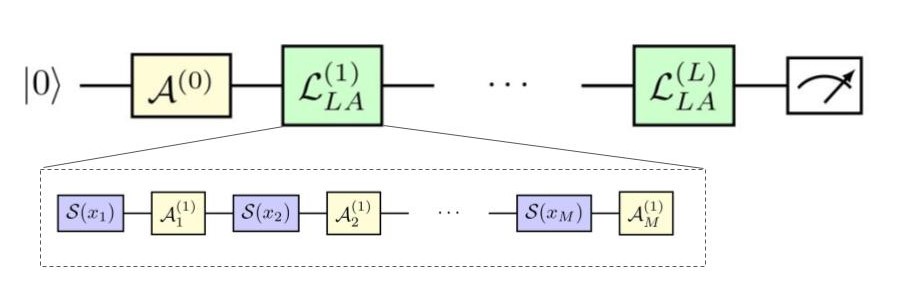} 
  \caption{Linear Ansatz}
  \end{subfigure}
  \begin{subfigure}[b]{0.5\textwidth}

  \includegraphics[width=.99\linewidth]{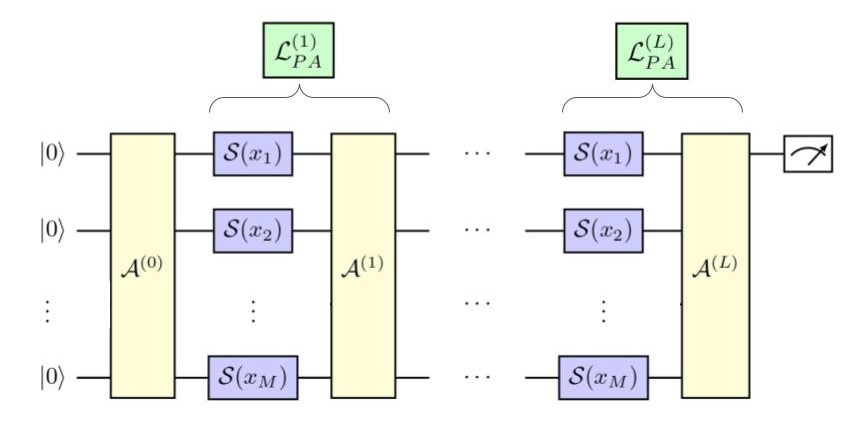} 
  \caption{Parallel Ansatz}
  \end{subfigure}
        \caption{Quantum circuit Ans\"atze of the QNN. The Linear Ansatz (a) encodes each data feature in a single qubit. Thus the circuit depth grows linearly with the total number of variables~$M$. The Parallel Ansatz (b) encodes the~$M$ variables in~$M$ qubits. In both Ans\"atze the circuit depth grows linearly with the number of layers $L$. Note that for $M=1$, both the Linear and Parallel Ansatz converge to the same linear circuit.}
        \label{fig:ansatzes}
\end{figure}

In this work, we implement a QNN to fit a 1D function $f (x)$ and obtain the Fourier series from the trained quantum circuit. 
Specifically, we train a single qubit circuit 
where we upload the data sampled from the function of interest.  
The circuit architecture is made of a linear
Ansatz that encodes each data feature, i.e. each of the $x$ values, in a single qubit, as shown in Fig.~\ref{fig:ansatzes}a, following the architecture introduced in~\cite{P_rez_Salinas_2020} given by an interleaved encoding circuit blocks and trainable circuit blocks.
The general form of the Ansatz is given by:
\begin{equation}
    U_0 \equiv \mathcal{A}(\vec{\theta_0})~, \qquad U_l \equiv \mathcal{A}(\vec{\theta_l})\mathcal{S}(x)~.
\end{equation}


More specifically, the circuit contains an initial operator to create a superposition state, $\mathcal{A}(\vec{\theta_0})$, then an encoding layer for the input data, $\mathcal{S}(x)$, and a following trainable layer to be optimized according to the adopted metrics, $\mathcal{A}(\vec{\theta_l})$. The cost function we adopt is given by the squared difference between the function output of the circuit (expected value in the computational base)~and the real value of the target function we want to fit.
With this approach, and as in~\cite{yu2022power}, the~expecta-tion value of this quantum model corresponds to a universal 1D Fourier series representation~\cite{Schuld_2021}:

\begin{equation}
    \langle \textit{M} \rangle (x, {\vec{\theta}}) = \sum_{w \in \Omega} c_w e^{ixw}~,
\end{equation}
where the elements of the Fourier series are given by the different elements in the quantum circuit: the encoding gate $\mathcal{S}(x)=e^{xH}$ gives the frequency $w$, under the assumption of the Hamiltonian $H$, which is $H=\frac{1}{2}\sigma_z$ for one qubit, being diagonal and the frequency spectrum is given by the sum of all different eigenvalues of $H$. It follows that the values of the Fourier coefficients, $c_w$, will be given by the weights $\vec{\theta_l}$ of the trainable unitary gates $\mathcal{A}(\vec{\theta_l)}$. 
Different data encoding strategies determine the degree of the output Fourier series which is the maximum frequency in the spectrum and is proportional to the number of layers $L$ in the circuit and the dimension of the computational space $D=max(\Omega)=(d-1)L$. For a one-qubit scheme: $d=2$. The higher the number of layers the more complex the function we can fit.

At this point, as done in \cite{Atchade-Adelomou:2023mjf}, using numerical methods to compute discrete Fourier transform, one can obtain, from the optimal weight of the parametrized circuit, the Fourier coefficients which represent the decomposition of our original function.
In Sec.~\ref{sec:qfiae}, we present a protocol to 
obtain the desired quantum integration representation.

In \cite{Casas:2023ure}, 
this approach has been extended to $M$-dimensional functions, where still the circuit depth grows linearly with the total number of layers $L$. Moreover, as shown in Fig.~\ref{fig:ansatzes}, the depth (Parallel Ansatz) or the width (Linear Ansatz) also grows linearly with the number of variables $M$. 
However, in this case, their proposed parallel Ansatz architecture represented in Fig.~\ref{fig:ansatzes}b cannot generate any multi-dimensional Fourier series like in the 1D case. Despite the large Hilbert space span by these unitaries, the total degrees of freedom (the complex number of Fourier coefficients) of the multi-dimensional Fourier series grow faster than the total available parameters in the quantum circuit. Nevertheless, in~\cite{Casas:2023ure} they show that this issue might be alleviated by extending the dimension of the special unitary group SU$(N)$, i.e. working with qudits instead of qubits. So this opens the possibility of creating a QNN for a qudit system with sufficient expressibility to fit multi-dimensional arbitrary functions.

From a general perspective, using more qubits allows for more expressive function families at shallower circuit depths, and indeed the number of qubits scaling as a function of data dimension is necessary for any potential quantum advantage (as quantum circuits of constant size are simulable in polynomial time in the depth). Indeed, the minimum is superlogarithmic scaling of the qubit numbers in the data dimension, and linear scaling can already ensure the exponential cost of the classical simulation of the quantum model using best-known classical algorithms.

This freedom also stymies any good approximations of how many qubits would be necessary to achieve good performance of quantum learning algorithms of this type, but at present it is not known how the increase in the number of qubits influences the quality of outcomes, and thus eventually outperform classical models.

\section{Quantum Fourier Iterative Amplitude Estimation}\label{sec:qfiae}

\begin{figure}[ht!]
       \centering
  \includegraphics[width=.991\linewidth]{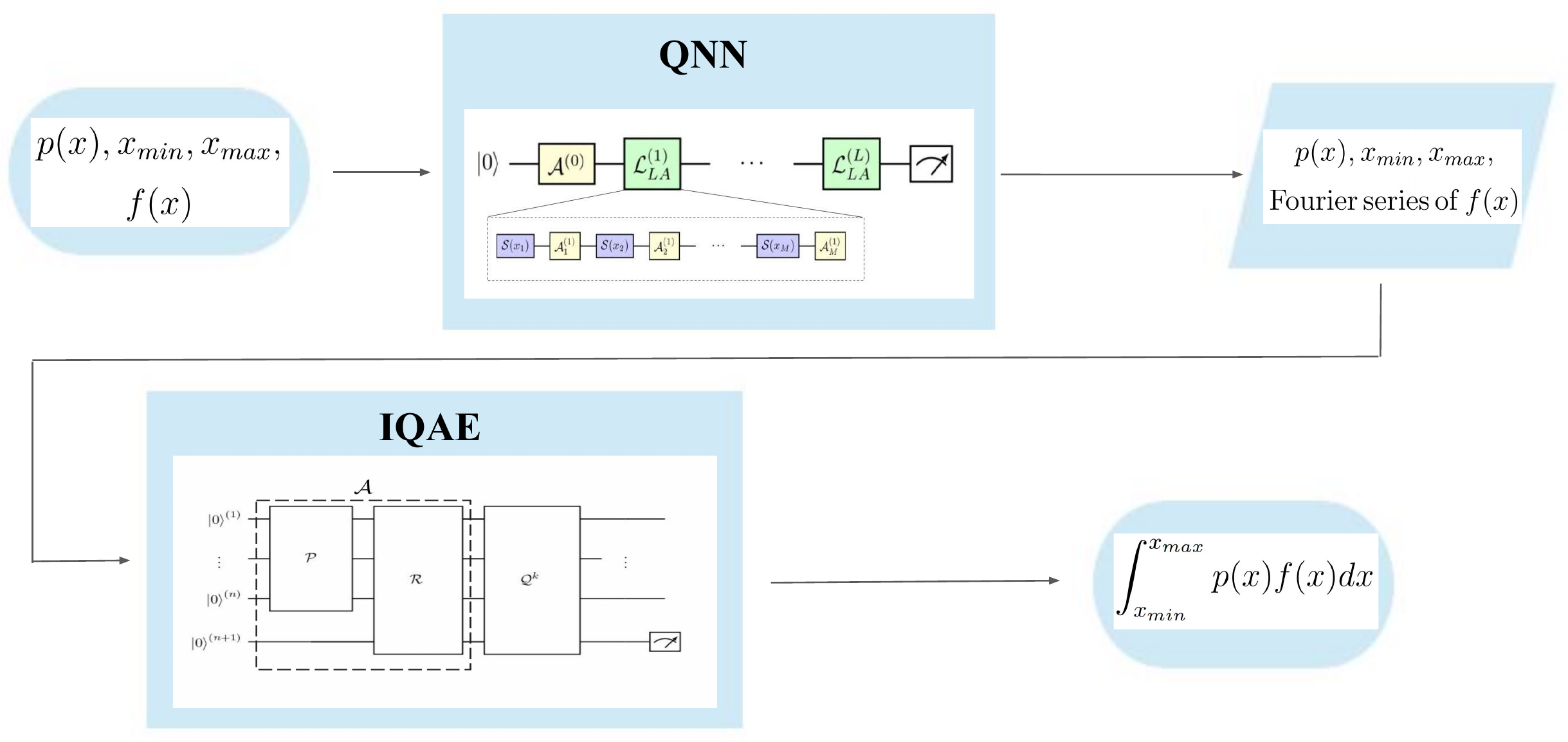}  

        \caption{Operational outline of QFIAE. Firstly, the input parameters including the function $f(x)$, the probability distribution $p(x)$, and the integration interval are introduced. Secondly, the QNN is employed to fit $f(x)$, and the Fourier series is obtained from the quantum circuit. Thirdly, $p(x)$, the integration interval, and the Fourier series of $f(x)$ are introduced as input parameters. Fourthly, IQAE calculates the integral for each trigonometric term in the Fourier series within the specified interval. Fifthly, the integrals are summed with their corresponding coefficients to derive the final integral result.  }
        \label{fig:sketch qfiae}
\end{figure}

In this section, we describe in detail with a test function our proposal for a quantum algorithm pipeline to compute integrals numerically. Our quantum integration algorithm is based on the quantum Monte Carlo integration method explained in Sec.~\ref{sec:montecarlo} combined with a QNN to obtain the Fourier coefficients of a function, as explained in Sec.~\ref{sec:qf}. We call this algorithm Quantum Fourier Iterative Amplitude Estimation~(QFIAE). In Fig.~\ref{fig:sketch qfiae}, a schematic diagram of the algorithm's operation is presented. 
\subsection{Benchmark example: integral of $1+x^2$}

As a benchmark example of the QFIAE method to calculate integrals the following integral is considered:
\beq
I=\int_0^1 \left(1+x^2\right) dx. \label{eq:integral_example}
\eeq

The motivation for this choice  comes from the fact that it is sufficiently streamlined to offer a pedagogical view of the functioning of the method for a well-known particle physics case. The integrand $f(x)=1+x^2$ represents a simple though non-trivial functional form of the differential cross-section of a scattering process: $e^+e^- \rightarrow q \bar{q}$, where no hadronic processes are considered to avoid the inclusion of parton densities.
In quantum electrodynamics (QED), this process is well defined by parametrizing the phase space with two angles, the total cross-section reads\footnote{Note that \Eq{eq:integral_example} and \Eq{eq:Xsection22} differ in the integration interval. Nevertheless, since the function is even, they are equivalent up to a factor of~$2$.}:
\begin{equation}
  \sigma \sim \int^{1}_{-1} \int^{2\pi}_0 \mathrm d \cos \theta \mathrm d \phi \left( 1+\cos^2 \theta\right) 
  \label{eq:Xsection22}
 \end{equation}
 
This integral has been computed using the IQAE algorithm within a different approach in~\cite{AGLIARDI2022137228}.
In that paper, a quantum Generative Adversarial Network (QGAN) is used to load the normalized distribution $1+x^2$.
To train the QGAN, classically generated samples are used, which allow the function $f(x)$ to be loaded into the quantum state.
Then, the integral is computed using IQAE. Although in~\cite{AGLIARDI2022137228} a quadratic speedup is achieved over the classical Monte Carlo simulation with respect to the number of queries, the total depth of the circuit given by the complexity of the QGAN architecture and the IQAE part makes this method to lie in the non-NISQ-friendly category. On the other hand, as discussed at the end of this section, the QFIAE method we propose presents a manageable depth for NISQ devices. Hence it is more likely to be successfully applied to in-real scenarios in the upcoming years. 

\begin{figure}[ht!]
       \centering
  \includegraphics[width=0.69\linewidth]{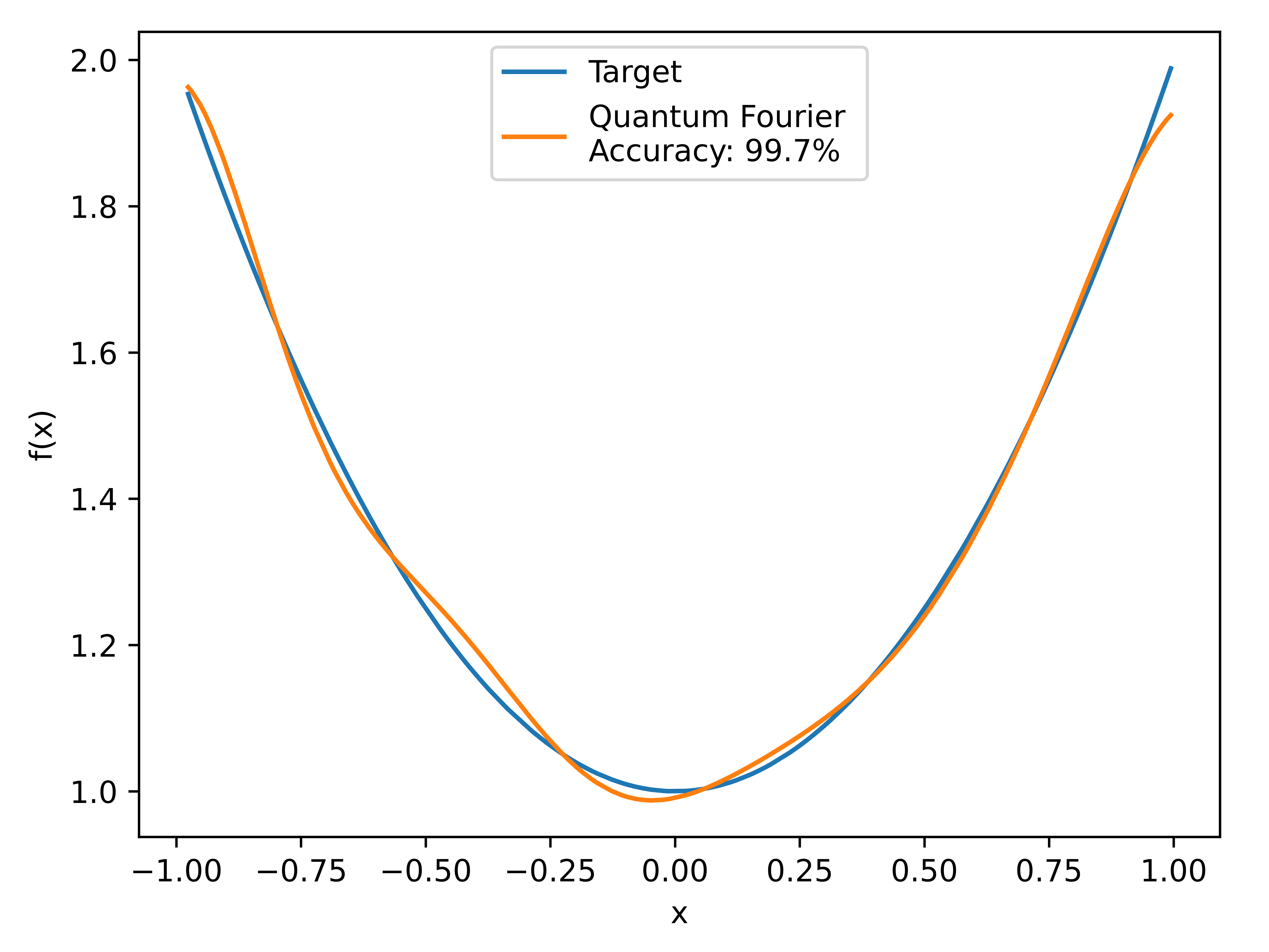}  

\caption{Quantum simulation using the Linear Ansatz in the QNN for fitting the target function $1+x^2$. We use $n_{qubits,\mathrm{QF}}=1$ and $L=10$, so the resultant Fourier series has $n_{Fourier}=10$ coefficients. A gradient descent method, the Adam optimizer \cite{kingma2014adam}, is utilized as a classical optimization subroutine with 200 data points in the range $[-1,1]$. The hyperparameters taken by the optimizer are \textit{learning\_rate}=0.05 and \textit{nepochs}=100.}
        \label{fig:qf_example}
\end{figure}

Starting from this reference example
we can test how our quantum algorithm succeeds in computing the same integral. Before applying the QFIAE method, the function $f(x)=1+x^2$ will be normalized since the QNN can only fit functions with an image from $[-1,1]$.
The first step of our quantum integration algorithm entails fitting the function through a QNN, as depicted in Fig. \ref{fig:qf_example}. The remarkable accuracy achieved by the QML model in fitting the function, exceeding 99\%, confirms the effectiveness of our method as a precise and reliable tool for fitting real functions. To optimize the model's performance, one can tune various hyperparameters, including the number of Fourier terms $n_{Fourier}$, training and test data size, and the number of iterations of the optimizer \textit{nepochs}. The values of these parameters are adjusted to decrease the model's learning time at the expense of accuracy or increase accuracy at the expense of longer learning time.

Hereafter, the Fourier coefficients are extracted from the quantum circuit and the resultant Fourier series is built. Let us explicitly see how this series will look like:
\beq
\begin{split}
f(x)& \approx 0.476 +1.169 \cos(x) -0.263\cos(2x) + \ldots  \\ & -0.017\cos(9x) + 0.004\cos(10x)+  \\ &
-0.125\sin(x)-0.278\sin(2x)+ \ldots  \\ &
-0.029 \sin(9x)-0.004\sin(10x).
\label{eq:fourierexample}
\end{split}
\eeq

After expressing the function as a Fourier series, as shown in~\Eq{eq:fourierexample}, the trigonometric components of the series are converted into the $\sin^2(ax+b)$ form and integrated over the desired interval using IQAE. In this example, the probability distribution that is considered is $p(x)=1/2^n$. This distribution is prepared as a shallow-depth quantum state so that the quadratic quantum advantage is maintained. In particular, it is generated by applying a $n$-dimensional Hadamard gate, 
$\mathcal{H}^{\otimes n}$.

Subsequently, the resulting integrals are summed with their corresponding coefficients to obtain an estimation of the integral presented in \Eq{eq:integral_example}. The results and comparison with the analytical value are shown in Table \ref{table:integral_example}.


 


\begin{table}[h!]
\begin{center}

\begin{tabular}{ |p{3.8cm} |p{2.3cm}|p{2.3cm}| }
 \hline
 \centering method = QFIAE & \centering $I_{est}$ &  \centering $\epsilon$  \cr 
 \hline
 \centering $n_{Fourier}=5$, $shots=100$ &
  \centering  \vspace{0.01cm} 1.32 $\pm$ 0.05 &    \centering \vspace{0.01cm} 0.991 \cr 
 \hline
 \centering $n_{Fourier}=10$, $shots=100$ &
  \centering  \vspace{0.01cm} 1.34 $\pm$ 0.06&    \centering \vspace{0.01cm} 1.006 \cr 
 \hline
 \centering $n_{Fourier}=5$, $shots=1000$ &
  \centering  \vspace{0.01cm} 1.33 $\pm$ 0.05&    \centering \vspace{0.01cm} 0.998 \cr 
 \hline
 
 \centering $n_{Fourier}=10$, $shots=1000$ &
  \centering  \vspace{0.01cm} 1.33 $\pm$ 0.04&    \centering \vspace{0.01cm} 0.999 \cr 
 \hline
\centering method = FQMCI & \centering $I_{est}$ &  \centering $\epsilon$  \cr 
 \hline
 
 \centering $n_{Fourier}=5$, $shots=100$ &
  \centering  \vspace{0.01cm} 1.35 $\pm$ 0.07&    \centering \vspace{0.01cm} 1.010 \cr 
 \hline
 \centering $n_{Fourier}=10$, $shots=100$ &
  \centering  \vspace{0.01cm} 1.34 $\pm$ 0.07&    \centering \vspace{0.01cm} 1.003 \cr 
 \hline
 \centering $n_{Fourier}=5$, $shots=1000$ &
  \centering  \vspace{0.01cm} 1.34 $\pm$ 0.07&    \centering \vspace{0.01cm} 1.007 \cr 
 \hline

 \centering $n_{Fourier}=10$, $shots=1000$ &
  \centering  \vspace{0.01cm} 1.34 $\pm$ 0.06&    \centering \vspace{0.01cm} 1.002 \cr 
 \hline

\end{tabular}
\caption{Integration of $1+x^2$ from $[0,1]$ using QFIAE and FQMCI. With $\epsilon=I_{est}/I_{exact}$.}
\label{table:integral_example}

\end{center}
\end{table}

In Table~\ref{table:integral_example}, an estimation of the integral of the function $1+x^2$ in the interval $[0,1]$ is presented. The proposed QFIAE method and the Fourier Quantum Monte Carlo Integration (FQMCI) method introduced in~\cite{Herbert_2022} are used to calculate the integral. The key difference between these two algorithms is how the Fourier series is calculated. In QFIAE, the Fourier series is estimated using a QNN, whereas in FQMCI, the Fourier series is computed classically through numerical integration. The integrals are computed for different values of the number of coefficients of the Fourier series, $n_{Fourier}$, and the number of shots of the IQAE algorithm, $shots$. In all the cases the IQAE algorithm employs a total of $n_{qubits,\mathrm{IQAE}}=4$ qubits, and the confidence interval and the estimated error are set to $\alpha=0.05$ and $\epsilon=0.01$ respectively. The third column of the table compares the obtained values with respect to the exact result ($I_{exact}=4/3$). Each integral in Table \ref{table:integral_example} is computed 50 times and then averaged to avoid statistical fluctuations and gain insight into which method performs better or which set of parameters works best. The errors are provided by the IQAE method of Qibo and could be reduced by adjusting the $\epsilon$ parameter. The analysis of the third column shows that QFIAE performs at least as well as FQMCI. Therefore, there is no loss of accuracy for estimating the Fourier series with a QNN. Furthermore, the best performance is obtained for $n_{Fourier}=10$ and $shots=1000$. This confirms the intuition that the greater the number of Fourier coefficients, the higher the expressibility of the Fourier series, and the greater the number of shots, the higher the accuracy.

\begin{table}[h!]
\begin{center}

\begin{tabular}{ |p{1.5cm}|p{1.5cm}|p{1.5cm}||p{1.5cm}|}
 \hline
 \multicolumn{4}{|c|}{${\rm QF}_{depth}=\mathcal{A}_{depth}+layers(\mathcal{A}_{depth}+\mathcal{S}_{depth})$} \\
 \hline
  \centering $\mathcal{A}_{depth}$& \centering $\mathcal{S}_{depth}$ &  \centering $layers$ &  \centering ${\rm  \,QF}_{depth}$ \cr 
 \hline
  \centering 3&  \centering 1 &  \centering 10 & \centering 43 \cr 
 \hline
 \multicolumn{4}{|c|}{${\rm IQAE}_{depth}=\mathcal{A}_{depth}+k\mathcal{Q}_{depth}$} \\
 \hline
   \centering $\mathcal{A}_{depth}$& \centering $\mathcal{Q}_{depth}$ &  \centering $k$ &  \centering ${\rm IQAE}_{depth}$ \cr 
 \hline
  \centering 4&  \centering 12 &  \centering 9\footnotemark{}& \centering 112 \cr 
 \hline

\end{tabular}

\caption{Depths of the Quantum Fourier (QF) and Iterative Quantum Amplitude Estimation (IQAE) parts of the QFIAE method.}
\label{table:depth}
\end{center}
\end{table}
\footnotetext{The value of $k$ is calculated by the algorithm and it changes in every iteration and integral. Hence the different values have been averaged, and the mean value is presented in the table.}

To assess the feasibility of our algorithm on current NISQ devices the Quantum Volume (QV) metric, introduced by IBM in~\cite{IBMQV}, might be considered. This metric quantifies the maximum size of square quantum circuits that can be successfully implemented by the computer. However, it should be noted that QV may not reliably predict the performance of deep and narrow circuits or wide and shallow circuits~\cite{ionqarticle}. This limitation is significant for our analysis, as Table~\ref{table:depth} shows two circuits that are much deeper than they are wide, with $n_{qubits, \mathrm{ QF}}=1$ and $n_{qubits,\mathrm{IQAE}}=4$. Therefore, the suitability of our circuits will be further evaluated by referring to a recent study conducted by IonQ \cite{ionqarticle}, which is presented in more detail in \cite{ionqnews}.
In their study, a collection of algorithms for quantum computers was tested and compared using QED-C benchmarks. The results indicate that the algorithm we present, with a quantum depth of 112 and a low number of qubits ($\leq 4$), can be executed with a high probability of success on IonQ and Quantinuum devices. This suggests that executing our proposed algorithm on current NISQ processors, at least in the trapped-ion devices of the mentioned companies, is not only feasible but also a promising endeavor.
\section{Conclusions and future work}
This paper proposes a new application of QML and QAE to build a novel Quantum Monte Carlo integrator, called QFIAE. This method goes beyond previously defined quantum integrators such as FQMCI, since it exploits QNN to approximate the Fourier series. Furthermore, since it avoids the necessity of numerical integration to obtain the Fourier coefficients for a general function $f(x)$, it might retain the quadratic quantum speedup that is achieved using amplitude amplification, i.e. Grover's algorithm, where other methods fail. This constitutes, to our knowledge, the very first end-to-end Quantum Monte Carlo integrator that might succeed in maintaining the speedup.

Moreover, we have successfully applied the quantum integrator to a one-dimensional function that corresponds to the elementary scattering process $e^+e^- \rightarrow q \bar{q}$. We obtain a very precise estimation of the integral with a relative error around~1\%, while maintaining a relatively low quantum depth. In particular, a study of the performance and the quantum depth of both circuits of the algorithm has been conducted and have been compared with the depth and width of the algorithms that can be effectively implemented on NISQ devices. The results depicted that, according to the feasibility discussion in Sec.~\ref{sec:qfiae}, for the ion-trapped devices of IonQ and Quantinuum our method lies in the region where a quantum algorithm can be executed with a high probability of success. 

The success of implementing a fully quantum Monte Carlo integrator for a one-dimensional function has potential consequences for future works.
On the one hand, further studies are necessary to evaluate the method's performance for $d$-dimensional functions. However, as commen-ted in Sec.~\ref{sec:qf}, the QNN for fitting multi-dimensional functions may require exploring even larger Hilbert spaces through the use of qudits. Thus the potential implementations on real devices may be limited.

On the other hand, there exist numerous potential applications where the implementation of the QFIAE method may prove advantageous. Particularly, high-energy physics, where complex integrals are commonplace, stands out as one of the fields that could significantly benefit from the development of an efficient quantum integrator. One area of noteworthy interest is loop Feynman integrals~\cite{deLejarza:2023XXX} since traditional methods can be computationally intensive, making them a challenging task. Moreover, they hold promise as suitable candidates for testing the effectiveness of QFIAE on more complicated functions.  

In conclusion, QFIAE is a promising end-to-end quantum algorithm for Monte Carlo integration that combines the power of PQC with Fourier analysis and IQAE to offer a new approach for efficiently approximating integrals with high accuracy. Although further research is necessary to explore its potential for more complex integrals and larger problem sizes, our results suggest that it is a valuable addition to the growing body of quantum integration methods.

\section*{Code}
A comprehensive tutorial of the Quantum Fourier Iterative Amplitude Estimation method using Qibo \cite{qibo_paper} can be found at: \url{https://qibo.science/qibo/stable/code-examples/tutorials/qfiae/qfiae_demo.html} \cite{tutorial_qfiae}.

\section*{Acknowledgment}

This work is supported by the Spanish Government (Agencia Estatal de Investigaci\'on MCIN/AEI/ 10.13039/501100011033) Grant No. PID2020-114473GB-I00, and Generalitat Valenciana Grants No. PROMETEO/2021/071 and ASFAE/2022/009 (Planes Complementarios de I+D+i, Next Generation EU). It also has been founded by the  Ministry of Economic Affairs and Digital~Transfor-mation of the Spanish Government through the QUANTUM ENIA project call - QUANTUM SPAIN project, and by the European Union through the Recovery, Transformation and Resilience Plan - NextGenerationEU within the framework of the Digital Spain 2026 Agenda, and by the CSIC Interdisciplinary Thematic Platform (PTI+) on Quantum Technologies (PTI-QTEP+). 
JML acknowledges financial support from Generalitat Valenciana (Grant No. ACIF/2021/219). LC is supported by Generalitat Valenciana GenT Excellence Programme (CIDE\-GENT/2020/011) and ILINK22045. MG is supported by CERN through the CERN QTI.

\bibliographystyle{JHEP}
\bibliography{bibliography}
\end{document}